\documentclass[3p, Garamond, 12pt]{elsarticle}

\usepackage{amssymb,color}
\usepackage{setspace}
\usepackage{amsthm,amsmath}

\usepackage[figuresright]{rotating}

\begin{document}
\doublespacing
\begin{frontmatter}




\title{Population growth, interest rate, and housing tax in the transitional China}

\author{Ling-Yun HE $^{a,b,*,}$\footnote{Dr. HE is a full professor of applied economics. WEN is a research associate supervised by Dr. HE.  The authors contribute equally in the project. The authors would like to thank our colleagues from JiNan University, Nanjing University of Information Science and Technology, and Peking University, for all their warm helps, constructive suggestions and pertinent comments. This project is supported by the National Natural Science Foundation of China (Grant Nos. 71273261 and 71573258).}  ~and Xing-Chun WEN $^{c}$     \\ \small a. School of Economics, JiNan University, Guangzhou 510632, China\\ \small b. School of Economics and Management, Nanjing University of Information Science and Technology, Nanjing 210044, China\\ \small c. Department of Applied Economics, Guanghua School of Management, Peking University, Beijing 100871, China\\ \small * Corresponding author.\\ \small Email: lyhe@amss.ac.cn}

\date{\small \emph{Submitted on \today}}

\begin{abstract}
This paper combines and develops the models in Lastrapes (2002) and Mankiw \& Weil (1989), which enables us to analyze the effects of interest rate and population growth shocks on housing price in one integrated framework. Based on this model, we carry out policy simulations to examine whether the housing (stock or flow) tax reduces the housing price fluctuations caused by interest rate or population growth shocks. Simulation results imply that the choice of housing tax tools depends on the kind of shock that housing market faces. In the situation where the housing price volatility is caused by the population growth shock, the flow tax can reduce the volatility of housing price while the stock tax makes no difference to it. If the shock is resulting from the interest rate, the policy maker should not impose any kind of the housing taxes. Furthermore, the effect of one kind of the housing tax can be strengthened by that of the other type of housing tax.
\end{abstract}

\end{frontmatter}


\section{Introduction}\label{intro}
At the end of 1999, the Chinese government basically canceled the housing benefit distribution system (\emph{Fu-Li-Fen-Fang} in Chinese), and then the market-oriented housing system began to take shape, eventually leading to the boom of the housing market. The Chinese people have suffered a lot from the continuously rising housing price. To curb the fast rising of housing price, on November 28, 2011, Shanghai and Chongqing in China became the first two that were officially chose to impose housing tax on individual house. However, the housing prices are still rising, which leads to the question: Do housing taxes work? There are many scholars focusing on this issue, but arriving at opposite conclusions. Some find that the housing taxes can help reducing the growth of housing price \cite{tiebout1956pure,oates1969effects,rosen1977note}, while others think that the housing tax policies lead to the rising of housing price \cite{Simon01051943,Hoyt2011}.

In our opinion, the different conclusions may result from the fact that there are sundry sources driving the housing price, such as money issues \cite{Swan199541,Lastrapes200240,Iacoviello2005,brunnermeier2008money,Wen2015257}, population factors \cite{Swan199541,Wen2015257,Mankiw1989235,Ohtake1996189}, etc. Since the present paper considers a China issue, we first summarize the main sources that drive the house price dynamics in China from the existing literature and then take the main factors into account in one framework.

There are various explanations for house price movements in China, among which population factors and monetary policy shocks are widely considered. The former attributes house price movements to changes in population, which are mostly fueled by the changing number of men in the marriage market \cite[e.g.,][]{NBERw18000}, and rural-urban migration and urbanization \cite[e.g.,][]{Chen20111}. In more detail, incorporating specific features of the Chinese economy, the way that the population affect the house price can be explained as follows. First, from the end of 1999, the market-oriented housing system in China began to take shape. After that, the changing population can affect the housing price dynamics through influencing the housing demand in the residential housing market. With this housing market reform, a huge demand for residential housing has been released, and as a consequence the housing price has increased a lot during the last decade. Second, the Chinese parents with a son want to pursue a bigger and more expensive house/apartment in a competitive manner to improve their son's relative attractiveness for marriage, which is a popular phenomenon in current Chinese society. As a result, the increasing number of men who are in marriage ages makes a large effect on the housing prices. Wei et al. \cite{NBERw18000} assume that housing is a status good in the Chinese marriage market and test its consequences for house prices. This work finds empirical evidence to support this hypothesis, and further an increase in the sex ratio resulted in 30-48 percents of the rise in urban house prices in China during 2003-2009. Third, big cities in China (like Beijing, Shanghai, and Guangzhou) receive a large number of migrants from inland provinces due to their rapid economic growth and employment opportunities and hence face greater pressure in the urban housing demand. Rapid urbanization process in China causes a dramatic rising in the urban population, which also leads to more housing demand. Chen et al. \cite{Chen20111} examine the possible effects of rural-urban migration and urbanization on housing prices of the cities in China, and find that the different urbanization levels and the migration situations have significant effects on the Chinese urban house prices. In conclusion, population factor related to market reform, marriage, and migration/urbanization is one of the most important factors in explaining the dynamics of housing prices in China.

As for the role of monetary policy shocks, many scholars argue that China's monetary policy actions are the key driving forces behind the increase of housing price in China. Using a non-linear modeling approach, Zhang et al. \cite{Zhang20122349} study the sources of housing prices movements in China, and find that monetary policies may be the key factors influencing house prices in China. Xu and Chen \cite{Xu201262} examine the impact of key monetary policy variables, including long-term benchmark bank loan rate, money supply growth, and mortgage credit policy indicator, on the real estate price growth dynamics in China. Empirical results consistently demonstrate that lower interest rate and loosening mortgage down payment requirement can help to accelerate the growth of housing price, and vice versa. Zhang \cite{Zhang201375} explores the causal relationship among the interest rate, house price inflation and consumer price inflation in China since 1998 through a standard multivariate dynamic model, and states that the recent housing market boom results from the low interest rate in China. These papers all suggest that the monetary policy actions of the Chinese government are the key drivers behind the fluctuations of the house price growth in China.

Therefore, for different sources that drive the housing price, will the effects of housing tax tools be different? In order to answer this, we combine and develop models in Lastrapes \cite{Lastrapes200240} and Mankiw and Weil \cite{Mankiw1989235} to consider both the interest rate and population growth in one framework, and take the housing stock and flow taxes into consideration.\footnote{Lastrapes \cite{Lastrapes200240}'s work considers the effect of money supply on housing price, while Mankiw and Weil \cite{Mankiw1989235}'s model only takes population growth into account.} Intuitively, if the housing boom is caused by the loose monetary policy, then the recently adopted policy packages aimed at the housing market should be focused on monetary policy tools, and in consequence the tax instruments may not be good measures to stabilize the house market. However, in the case where the driving force of the housing market is the population factor, the housing tax may slow down the rising of the housing market as the household would choose to rent instead of to buy a new house or delay the demand for the house. As a result, the effect of the housing taxes may depend on the driving force of the housing market.

The main contributions of our work are as follows: we propose a theoretical model and carry out simulations to answer the questions mentioned above, which can also be used for other policy analysis; we conclude that the effect of housing tax depends on what shock (interest rate or population growth shock) the housing market is affected by, which will provide new insight into the current research on housing tax in China. In what follows, Section 2 sets up the model, Section 3 presents the simulation results, and Section 4 concludes.

\section{The Model}
In this paper, we propose a partial housing market equilibrium model, which includes housing demand and supply equations. The housing demand equation is derived follow the work of Lastrapes \cite{Lastrapes200240}, and the supply equation draws the model in Mankiw and Weil \cite{Mankiw1989235}. As a combination, our model takes the population growth and interest rate into account.

Let $N_{t}$ be the population in the economy, $H_{t}$ be the aggregate stock of housing, and $q_{t}$ be the real housing price. First, we derive the housing demand equation. Assume all the people are homogeneous, and face the same choice. Let an aggregate consumer represents the whole population, and maximize the intertemporal objective function:
\begin{equation}\label{eq.1}
V_{0}=\sum_{t=0}^{\infty}\beta^{t}U(C_{t},H_{t})
\end{equation}
\noindent where $C_{t}$ is non-housing aggregate consumption, and $\beta$ is the discount factor. The consumer maximizes its utility subject to:
\begin{equation}\label{eq.2}
 A_{t+1}+C_{t}+q_{t}H_{t}[1+\tau_{s}+\tau_{f}]+\frac{1+R_{m,t}}{1+\pi_{t}}B_{m,t-1}
 =y_{t}(1-\tau)+\frac{1+R_{t}}{1+\pi_{t}}A_{t}+(1-\delta+\tau_{f})q_{t}H_{t-1} +B_{m,t}
\end{equation}
The left-hand side of this budget constraint (in real terms) contains the outflow of funds: the real value of non-mortgage financial assets carried over into the next period ($A_{t+1}$), consumption, purchases of housing stock ($q_{t}H_{t}$), and expenditures on mortgage loans ($R_{m,t}$ is the nominal yield on mortgage-secured loans). The right-hand side includes the inflow of funds: real income ($y_{t}$) after tax ($\tau$ is the income tax rate), the current holdings of non-mortgage financial assets ($R_{t}$ denotes the nominal interest rate on this), the value of the housing stock net of depreciation (at rate $\delta$), and new mortgage borrowing. $\pi_{t}=P_{t}/P_{t-1}-1$ denotes the inflation rate. $\tau_{s}$ and $\tau_{f}$ are taxes on housing stock and flow. The flow tax is defined by $\tau_{f}(q_{t}H_{t}-(1-\delta)q_{t}H_{t-1})$, while the stock tax is $\tau_{s}q_{t}H_{t}$. Following Lastrapes \cite{Lastrapes200240}, we impose the mortgage borrowing constrain:
\begin{equation}\label{eq.3}
 B_{m,t}=\gamma q_{t}H_{t}, \gamma\in(0,1)
\end{equation}
\noindent where $\gamma$ is the loan-to-value ratio.

In the following part, we will obtain the first order conditions in solving the model, and derive the housing demand equation. To simplify, we substitute \eqref{eq.3} into \eqref{eq.2} to eliminate $B_{m,t}$. Thus, the consumer chooses $C_{t}$, $A_{t+1}$, and $H_{t}$ to maximize \eqref{eq.1} subject to \eqref{eq.2}. Let $\lambda_{t}$ be the Lagrangian multiplier associated with the mortgage constraint \eqref{eq.3}. The first order conditions are:
\begin{equation}\label{eq.4}
\partial C_{t}: U_{C,t}=\lambda_{t}
\end{equation}
\begin{equation}\label{eq.5}
\partial A_{t+1}: \lambda_{t+1}\frac{1+R_{t+1}}{1+\pi_{t+1}}-\lambda_{t}=0
\end{equation}
\begin{equation}\label{eq.6}
\partial H_{t}: U_{H,t}-\lambda_{t}q_{t}(1+\tau_{f}+\tau_{s}-\gamma)-\lambda_{t+1}(\frac{1+R_{m,t+1}}{1+\pi_{t+1}}\gamma q_{t}+(1+\tau_{f})(1-\delta)q_{t+1})=0
\end{equation}
\noindent where $U_{x,t}$ is the partial derivative of $U$ with respect to $x$ ($x=C,H$). Substituting \eqref{eq.4} and \eqref{eq.5} into \eqref{eq.6} to eliminate the multipliers and making some obvious approximations for the ratios of interest rates, then obtain
\begin{equation}\label{eq.7}
\frac{U_{C,t}}{U_{H,t}}=q_{t}[\theta+(1+R_{m,t+1}-R_{t+1})\gamma]-q_{t+1}[(1+\tau_{f})(1-\delta)-R_{t+1}+\pi_{t+1}]
\end{equation}
\noindent where $\theta=1-\gamma+\tau_{f}+\tau_{s}$. Set the utility function $U$ in a Cobb-Douglas fashion:
\begin{equation}\label{eq.8}
U(H_{t},C_{t})=\alpha \log(C_{t})+(1-\alpha) \log(H_{t})
\end{equation}
Besides, divide $C_{t}$ and $H_{t}$ by $N_{t}$. Then rewrite \eqref{eq.7} as
\begin{equation}\label{eq.9}
(\frac{1-\alpha}{\alpha})\frac{c_{t}}{h_{t}}=q_{t}[\theta+(1+R_{m,t+1}-R_{t+1})\gamma]-q_{t+1}((1+\tau_{f})(1-\delta)-R_{t+1}+\pi_{t+1})
\end{equation}
\noindent where $c_{t}=C_{t}/N_{t}$ and $h_{t}=H_{t}/N_{t}$. We assume that $\log(c_{t})$ is constant, then take logs on \eqref{eq.9} to solve for $q_{t}$. Then, we apply the first order Taylor approximation on the log functions to obtain the log-linear housing demand equation (for more details, see the appendix in Lastrapes \cite{Lastrapes200240}):
\begin{equation}\label{eq.10}
\log(q_{t})=K_{1}-w_{1}\log(h_{t})+w_{2}\log(q_{t+1})+w_{3}(R_{t+1}-R_{m,t+1})-w_{2}(R_{t+1}-\pi_{t+1})
\end{equation}
\noindent where $K_{1}$ collects all constants, and $r=R-\pi$. The values of $w_{1}$, $w_{2}$, and $w_{3}$ are determined by the steady state of the housing market equilibrium, which are given as
\begin{equation}\label{eq.w}
w_{1}=\frac{(\frac{1-\alpha}{\alpha})\frac{c}{h}}{(\frac{1-\alpha}{\alpha})\frac{c}{h}+((1-\delta)(1+\tau_{f})-r)q}, ~~~~w_{2}=1-w_{1},~~~~w_{3}=\frac{(1+R_{m}-R)\beta}{(1+R_{m}-R)\beta+\theta}
\end{equation}
where we drop the $t$ subscript to denote the steady-state value of a particular variable. The equation \eqref{eq.10} defines the housing demand.

Next, we define the supply equation following the work of Mankiw and Weil \cite{Mankiw1989235}. Assume that gross investment in housing is taken to be an increasing function of the real housing price $q_{t}$ and proportional to the scale of the economy as measured by the population $N_{t}$:
\begin{equation}\label{eq.11}
\Delta H_{t}=\psi(q_{t})N_{t}-\delta H_{t}, \psi'>0
\end{equation}
Let $n_{t}$ be the rate of growth of the population $N_{t}$, $n_{t}=\frac{\Delta N_{t}}{N_{t}}$. To simplify, we set $\psi(q_{t})=\kappa q_{t}$. Then, we rewrite \eqref{eq.11} terms of $h_{t}$. Differentiating $H_{t}/N_{t}$ with respect to time and substituting gives
\begin{equation}\label{eq.12}
h_{t+1}-h_{t}=\kappa q_{t}-(n+\delta)h_{t}
\end{equation}
Thus, the population growth $n_{t}$ is added into the housing model. Equations \eqref{eq.10} and \eqref{eq.12} determine the partial equilibrium dynamics of the housing market.

\section{Simulation Results}
In this section, we calibrate the parameters to the real data in China and stimulate the model to simulate the dynamics of price response. By simulating the theoretical model, we can quantitatively assess the effects of housing taxes on the responses of real housing price to interest rate and population shocks.

\subsection{Parameter Calibration}
To perform the simulation, we solve the steady state of the housing market equilibrium using the parameter values set to match the economic data from China.\footnote{We carry out the simulations by Dynare software with Matlab. The parameters are obtained from the China's economic data.} We set the values of $q$, $h$, $n$, and $\pi$ to be 1, 1, 1\%, and 3\%, respectively. The steady state (annual) nominal interest rate $R$ and $R_{m}$ are 5\% and 8\%. The ratio $\frac{c}{h}$ is set to be 0.267. $\alpha$, $\gamma$, and $\delta$ are set at 0.85, 0.8, and 0.02, respectively. $\kappa$ is decided by equation \eqref{eq.12} in steady state. The exogenous shocks are assumed to follow AR(1) processes\footnote{As we can see, the shock to interest rate is negative. The reason for this is that we want to produce a positive response of housing price to interest rate shocks. As we know, a decrease in the interest rate leads to an increase in the housing price, which is also tested in the simulation results.}, which are described as:
\begin{equation}\label{eq.13}
\ln(R_{t})=(1-\rho_{R})\ln(R)+\rho_{R}\ln(R_{t-1})-e_{R,t}
\end{equation}
\begin{equation}\label{eq.14}
\ln(n_{t})=(1-\rho_{n})\ln(n)+\rho_{n}\ln(n_{t-1})+e_{n,t}
\end{equation}
where $e_{R,t}$ and $e_{n,t}$ are  independent and identically distributed (i.i.d.) processes with constant variances $\sigma_{R}^2$ and $\sigma_{n}^2$. In the simulations, we set $\rho_{R}=\rho_{n}=0.8$ and $\sigma_{R}^2=\sigma_{n}^2=0.01$. At each round, we change the values of $\tau_{s}$ and $\tau_{f}$, and calculate the values of $K_{1}$, $w_{1}$, $w_{2}$ and $w_{3}$ according to their definitions. Table \ref{tab.1} summarizes the values and definitions of the parameters used for simulation.

\begin{table}[htbp]
\centering
\small
\caption{Calibrated Parameters}
\begin{tabular}{ccccccccccccccc}
\hline
Parameter & $q$ & $h$ & $n$ & $\pi$ & $R$ & $R_{m}$ & $c/h$ & $\alpha$ & $\gamma$ & $\delta$ & $\kappa$ & $K_{1}$ & $w_{1}$, $w_{2}$, $w_{3}$\\
\hline
Description & 1 &   1 &  1\% &  3\% &  5\% & 8\%    & 0.267 & 0.85    &      0.8  &  0.02   & Eq. \eqref{eq.12} & Eq. \eqref{eq.10} & Eq. \eqref{eq.w} \\
\hline
\end{tabular}%
\label{tab.1}%
\end{table}

\subsection{Results}
\begin{figure}[htbp]
\centering
\includegraphics[width=0.7\textwidth]{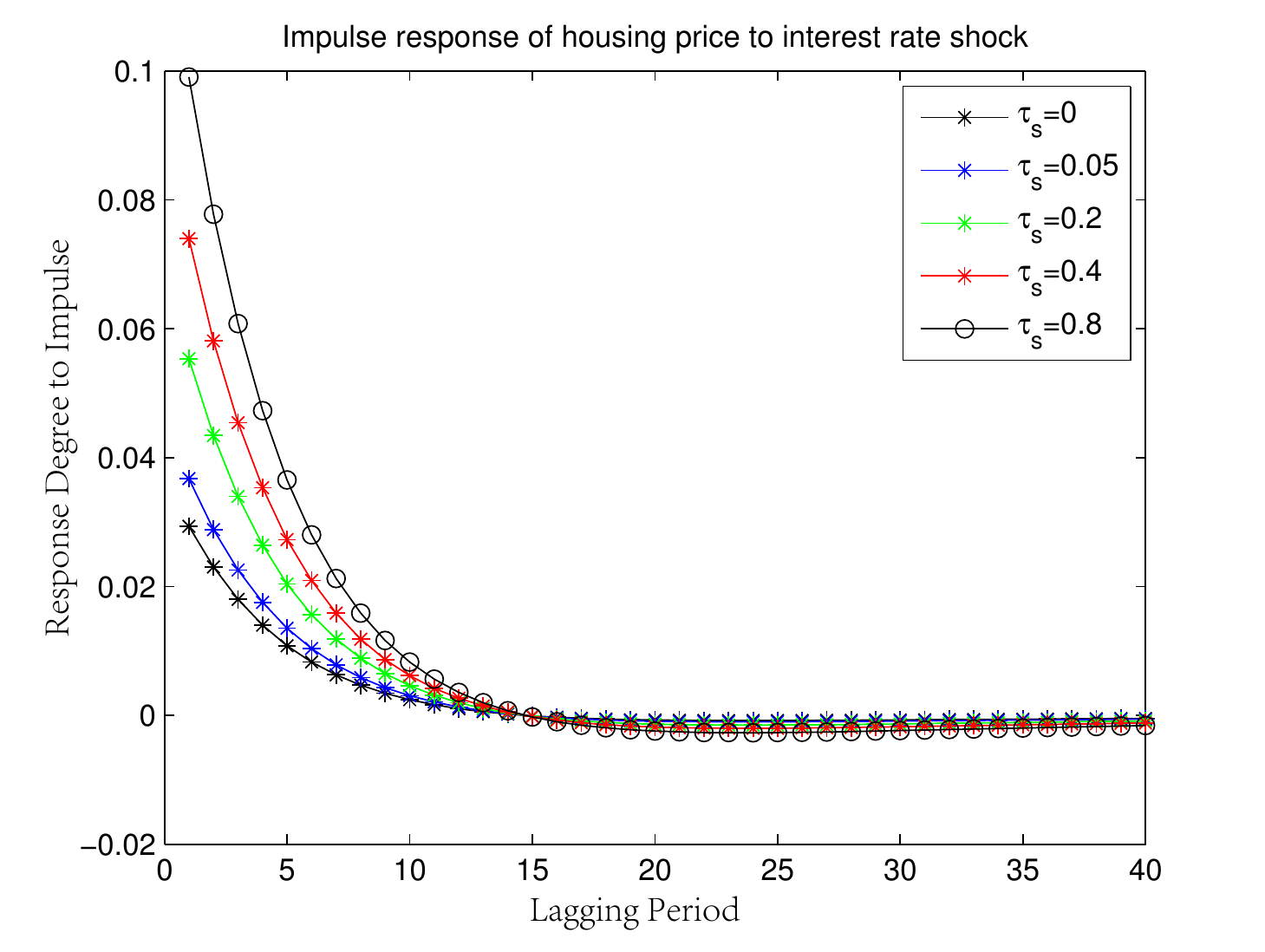}
\caption{Responses of housing price to interest rate shock (different $\tau_{s}$)} \label{fig.1}
\end{figure}

\begin{figure}[htbp]
\centering
\includegraphics[width=0.7\textwidth]{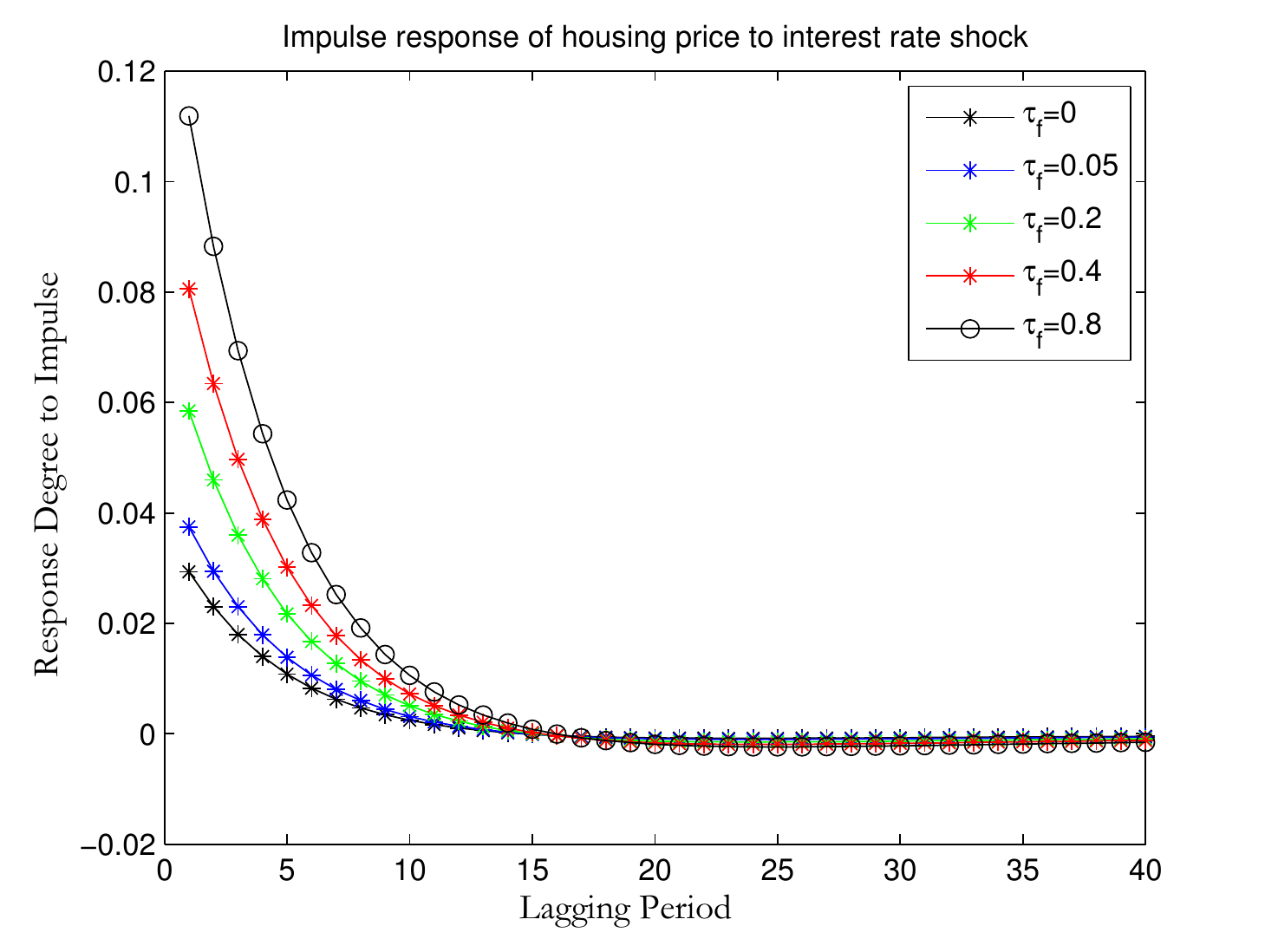}
\caption{Responses of housing price to interest rate shock (different $\tau_{f}$)} \label{fig.2}
\end{figure}

\begin{figure}[htbp]
\centering
\includegraphics[width=0.7\textwidth]{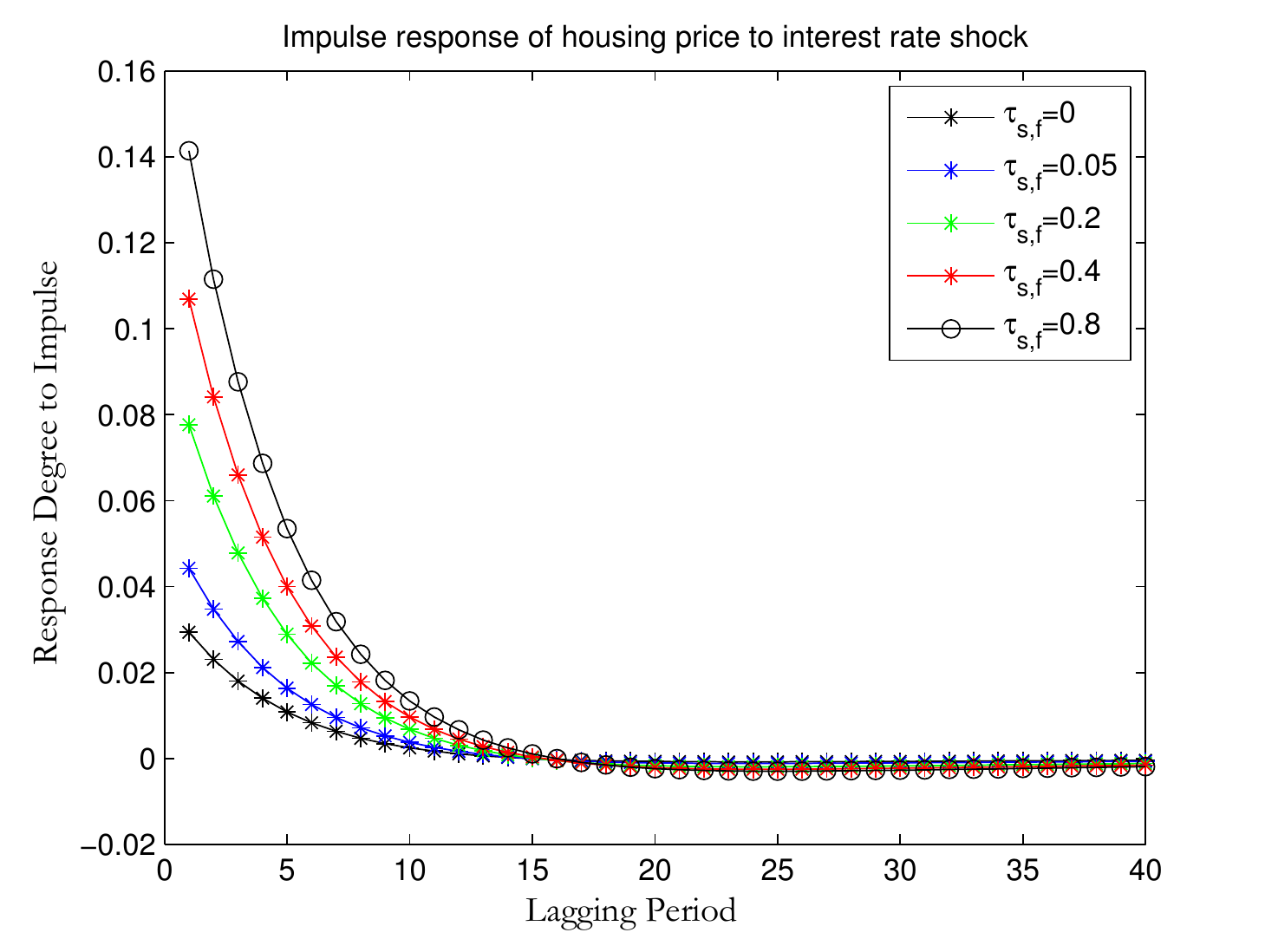}
\caption{Responses of housing price to interest rate shock (different $\tau_{s,f}$)} \label{fig.3}
\end{figure}

\begin{figure}[htbp]
\centering
\includegraphics[width=0.7\textwidth]{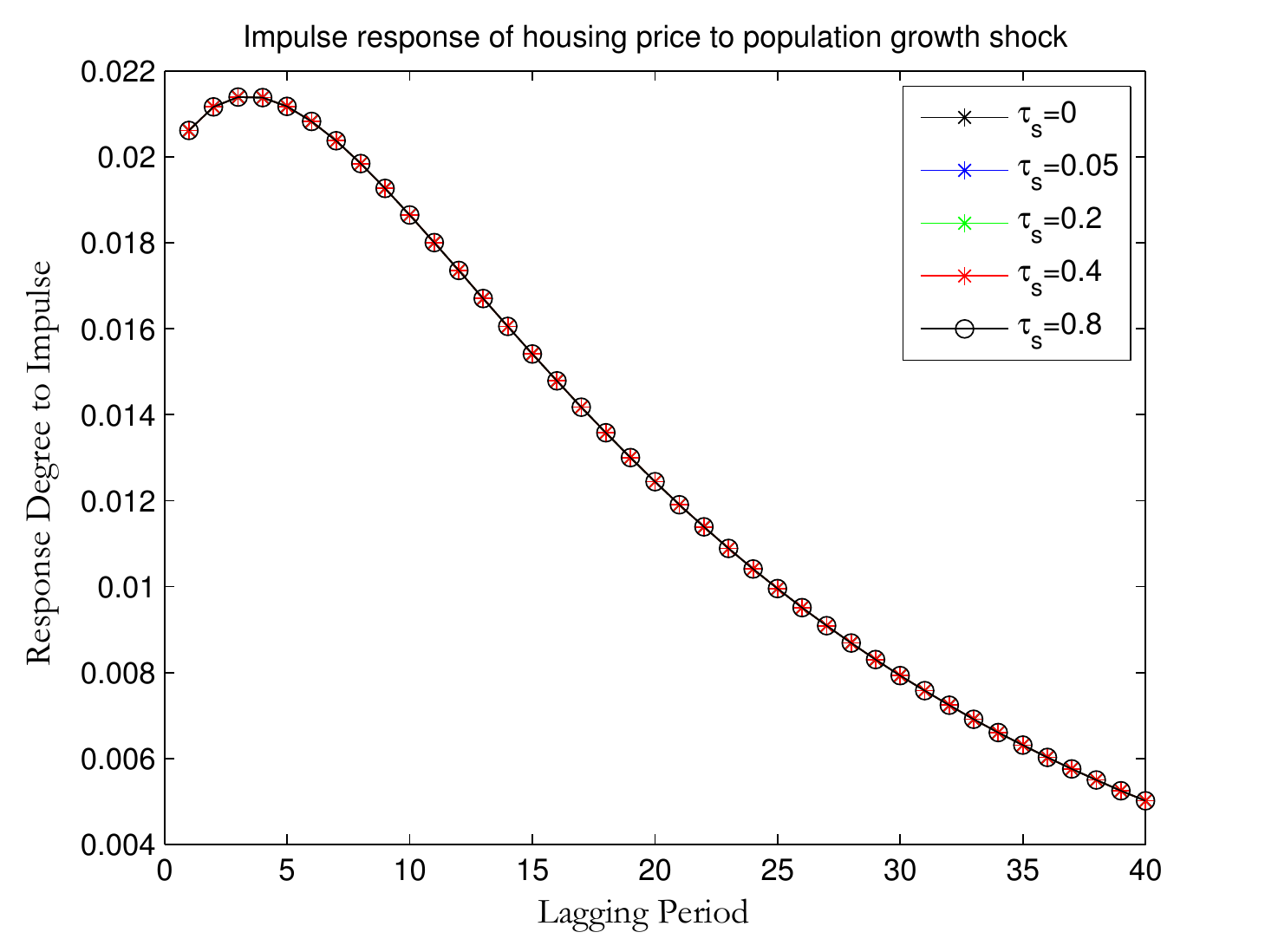}
\caption{Responses of housing price to population growth shock (different $\tau_{s}$)} \label{fig.4}
\end{figure}

\begin{figure}[htbp]
\centering
\includegraphics[width=0.7\textwidth]{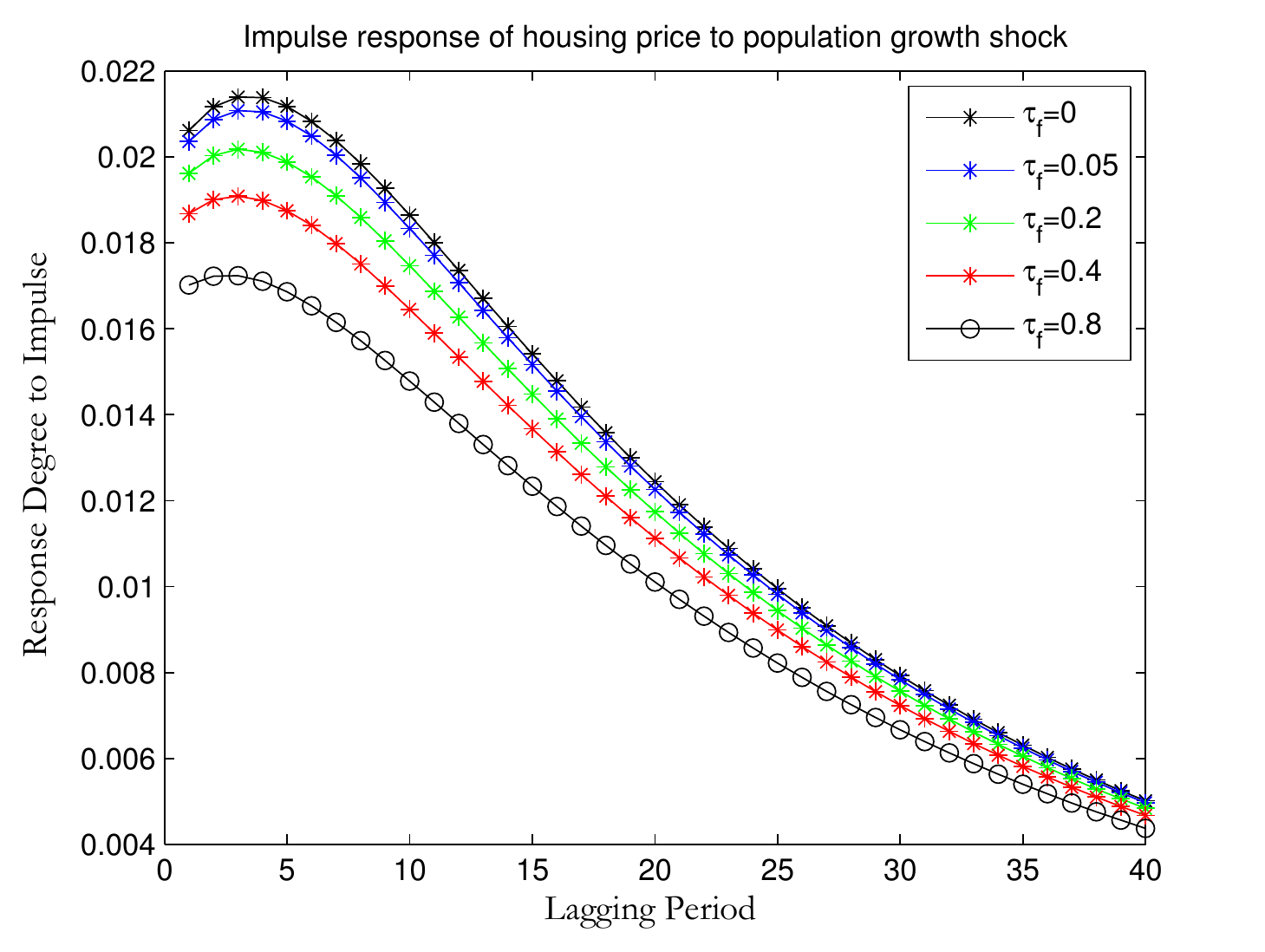}
\caption{Responses of housing price to population growth shock (different $\tau_{f}$)} \label{fig.5}
\end{figure}

\begin{figure}[htbp]
\centering
\includegraphics[width=0.7\textwidth]{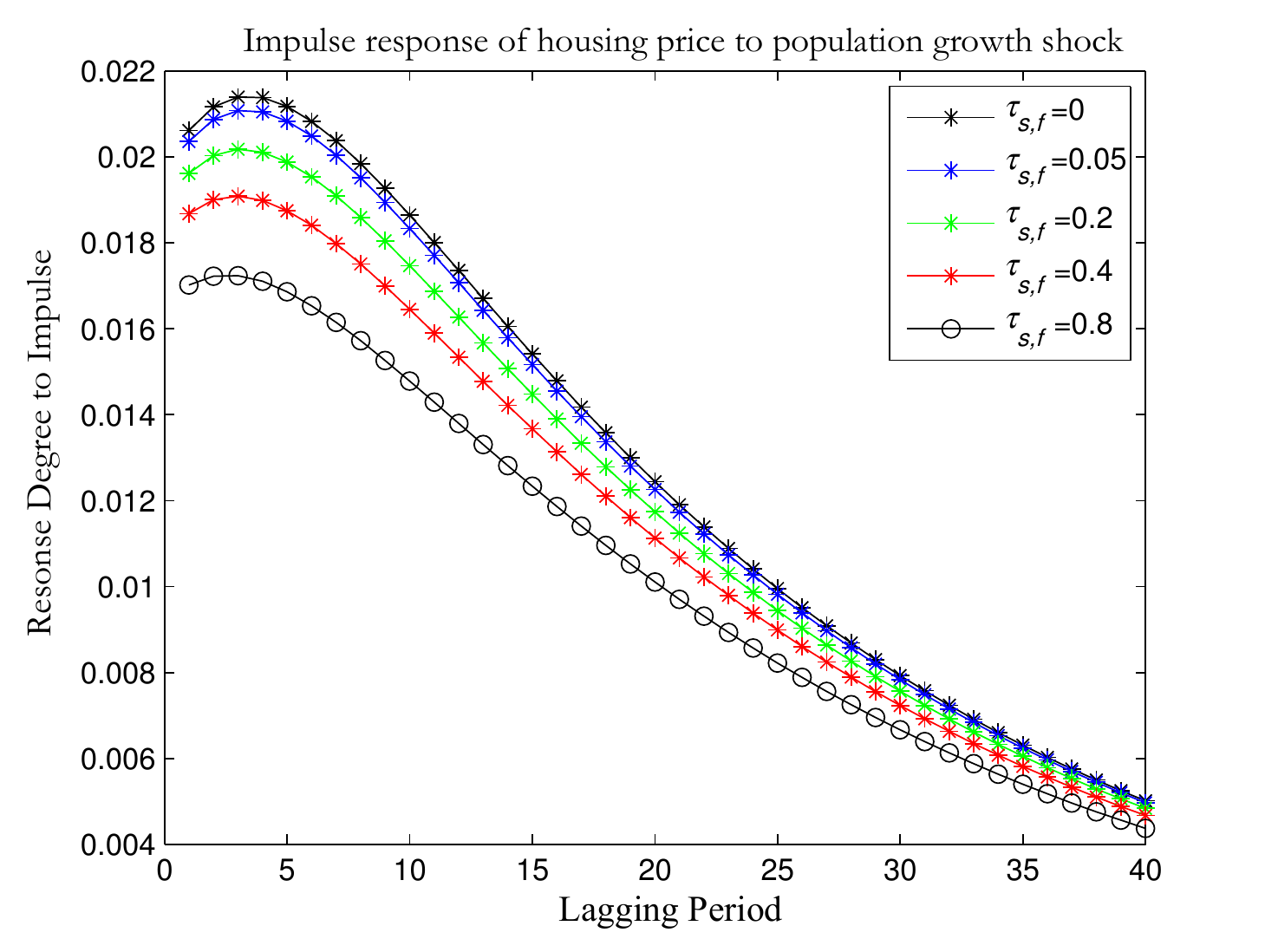}
\caption{Responses of housing price to population growth shock (different $\tau_{s,f}$)} \label{fig.6}
\end{figure}

Figures \ref{fig.1}-\ref{fig.6} represent the simulation results. As we can see from these figures, a negative shock to interest rate increases the housing price, which matches the empirical results from the literature. A rising population means an increasing demand for housing. As a consequence, the positive shocks lead to a positive response of housing price in China.

Turning to the effects of the housing taxes, from Figures \ref{fig.1} and \ref{fig.2}, we can see that raising the rates of flow or stock tax causes more housing price volatilities when facing the interest rate shock, which implies that both the tax tools are making things worse rather than improving the situation. 

Worse results can be found in Figure \ref{fig.3} if we take the two tax tools together. The results show that both types of housing taxes contribute to generating more volatilities of housing price, and that their effects are mutually reinforcing. This implies that housing tax instruments are not proper ways to regulate the housing market if the housing market boom is driven by the low interest rate.

However, if housing prices are affected by the population growth shocks, the results are quite different. Figure \ref{fig.4} shows that changing the rate of stock tax has no effect on the housing price. But in Figure \ref{fig.5}, a higher housing flow tax rate reduces volatility of the housing price. 

We also find the same result in Figure \ref{fig.6}, where we implement both tax measures. Similarly, the effects of the two housing tax instruments can be mutually reinforcing. As a result, in the case where the housing price boom is caused by the positive population shock, raising the rate of flow tax is a good choice to stabilize the housing market.

Therefore, there are two main findings: First, facing the shock of interest rate, applying the stock or flow housing tax is just to make the situation worse, namely, both the tax tools lead to bigger response of housing price. Second, the flow housing tax is effective in the housing market facing positive shock from the population growth.

\section{Conclusions}
From the existing literature, interest rate and population growth shocks are two key drivers of the housing price movements in China. In this paper, we propose a theoretical model that considers both the population and interest rate shocks in one framework, and incorporate the housing tax into this model. After that, we carry out numerical simulations. Our simulation results answer the two questions:
 
One, do housing taxes work? The answer is that the effect of housing tax depends on what shock the market is affected by. Different shocks need different policy measures. The policy makers should not depend on a single tool to regulate the housing market since there are different shocks influencing the housing market. As a result, the policy makers should not jump in with a quick and easy solution, without having spent time to first understand and then carefully analyze the sources of shocks in the housing market.

Two, will the effects of housing tax tools be different facing different driving forces of the housing price movements in China? The results suggest that only the flow tax works in the situation where the housing price volatility is caused by the population growth shock. If the shock comes from interest rate, it will make things worse if we conduct the housing taxes. Therefore, our work provides little support for Shanghai and Chongqing cities to impose house property taxes on individual housing. Without the critical understanding of the different sources of shocks,  such taxes would be the seemingly simple and easy solutions, but to the wrong problem.

\newpage
\section*{Reference}
\bibliographystyle{elsarticle-num}
\bibliography{ref1}

\end{document}